\documentclass{aa}
\usepackage{graphicx}
\usepackage{amssymb,amsmath}
\newcommand{\mdens}{{\rm g~cm^{-3}}}

\newcommand{\msun}{{\rm M}_\odot}

\def\drom{{\rm d}}

\def\rcore{R_{\rm core}}
\def\msuny{M_\odot~{\rm y}^{-1}}
\def\msuny9{10^{-9}~\msuny}

\def\muco{\mu_{\rm core}}

\usepackage{natbib}
\bibpunct{(}{)}{;}{a}{}{,} 
\usepackage{color}
\usepackage[dvipsnames]{xcolor}

\def\drom{{\rm d}}

\begin{document}

\title{Neutron star properties and the equation of state for its core}
\author{ J. L. Zdunik \and M. Fortin \and P. Haensel}
\institute{N. Copernicus Astronomical Center, Polish
           Academy of Sciences, Bartycka 18, PL-00-716 Warszawa, Poland
{\tt jlz@camk.edu.pl, fortin@camk.edu.pl, haensel@camk.edu.pl}}
\offprints{jlz@camk.edu.pl}
\date{Received xxx Accepted xxx}

\abstract{Few unified equations of state for neutron star matter where core and crust are described using the same nuclear 
model are available. However the use of non-unified equations of state with a simplified matching between the crust and 
the core has been shown to introduce uncertainties in the radius determination which can be larger than the expected precision 
of the next generation of  X-ray satellites. }
{We aim at eliminating the dependence of the radius and mass of neutron star 
on the detailed model for the crust and on the crust-core matching procedure. 
} 
{We solve the approximate equations of the hydrostatic equilibrium for the crust of neutron stars obtaining 
a precise formula for the radius which depends only 
on the core mass and radius, and on the baryon
chemical potential at the core-crust interface and on the crust surface. For a fully accreted crust one needs additionally 
the value of the total deep crustal heating per one accreted nucleon.}
{For typical neutron star masses the approximate approach allows to 
determine the neutron star radius with an error $\sim 0.1\%$ ($\sim 10\,$m, equivalent to a 1\% inaccuracy in 
the crust thickness).
The formalism applies to neutron stars with a catalyzed or a fully accreted crust. 
The difference in the neutron star radius between 
the two models is proportional to the total energy release due to deep crustal heating.}
{For a given model of dense matter describing the neutron star core, the radius of a neutron star
can be accurately determined independently of the crust model with a precision much better than the 
$\sim5\%$ one expected from the next generation of 
X-ray satellites. This allows to circumvent the problem of the radius uncertainty which may 
arise when non-unified equations of state for the crust and the core are used. } 


\titlerunning{Neutron star properties and the core equation of state}
\maketitle

\section{Introduction}
\label{sect:introduction}
The interior of a neutron star (NS) consists of two main parts: the liquid 
core and the solid crust. While the core is uniform (homogeneous),  the crust is non-uniform composed of nuclear clusters.
Consequently,  calculating the crust equation of state (EOS) is much less-straightforward 
than for the core, explaining the smaller number of crust EOS available compared to 
the ones for the core. 
In particular few unified EOS, i.e. based on the same nuclear model for the crust and core,
have been developed, 
see eg. \cite{DH,BSka,BSkb,Fortin2016}. Therefore non-unified EOS are often used, assuming different 
nuclear interaction models   for the crust and the core.
As shown in \cite{Fortin2016}, for masses of astrophysical interest ($M>1\,\msun$) the use of 
non-unified EOS can introduce an uncertainty on the radius determination of the order of $5\%$, 
as large as the precision expected from the next generation of X-ray telescopes: NICER 
\citep{NICER}, Athena \citep{Athena+} and potential LOFT-like missions
\citep{LOFT}.

The solid crust of a NS with a mass $M>1\,\msun$ contains only about one percent of star's 
total mass. However, the crust is believed to play an important role in many NS phenomena, e.g. pulsar glitches, X-ray bursts, gamma-ray 
flares of magnetars, torsional oscillations of NS, cooling of isolated NS, cooling of X-ray transients (for review see
e.g, \citealt{Chamel2008rev}). A standard model of NS crust assumes that it is built of matter in nuclear equilibrium, 
the thermal corrections are negligibly small, and at a given baryon number density $n$,  the crust matter is in 
a state of minimum energy per nucleon, $E$. In such a state which defines the ground state (GS) of matter, the matter
is called {\it catalyzed}.   As the density $n$ (or the mass-energy 
density $\rho$) can undergo discontinuous  jumps inside the NS, a more suitable independent variable is the pressure $P$, which is 
strictly monotonous within the star. The thermodynamic potential is then the Gibbs free energy 
(the baryon chemical potential) $\mu=E+P/n$ which replaces $E$.  In the strict GS of matter both $P$ and $\mu$ 
are continuous and monotonously increasing with the density when going towards the NS center. 

The GS approximation is expected to be good for isolated NS born in core-collapse supernovae. 
However, a significant fraction of NS remains for $10^8 - 10^9$~yr in low-mass X-ray binaries (LMXB), 
where they undergo a phase of accretion of matter from its evolved companion star. 
During  the LMXB stage the NS is spun-up to millisecond periods, this is the so-called pulsar recycling. 
In such accreting NS the original crust has been replaced (fully or partially) by an accreted one. 
In what follows we consider only a fully accreted crust, i.e. we assume that the NS has accreted matter 
with an integrated rest mass larger than the rest mass of the original GS crust. 

The composition of an accreted crust (AC) is expected to be very different from the 
one of a crust built of catalyzed matter in the GS. However, the neutron drip, dividing the whole 
crust into the outer  
(nuclei in electron gas) and the inner (nuclei in neutron gas and electron gas) crusts is found at similar 
 density for GS and AC cases. \cite{Chamel2015ndrip} find, using up-to-date energy density functionals and 
the Hartree-Fock-Bogoliubov method for solving the nuclear many-body problem, 
$\rho_{\rm ND}^{\rm (GS)}=4.3 - 4.4\times 10^{11}~\mdens$. For an accreted crust they find some 
dependence on the energy density functional model, as well as on the initial composition of ashes of X-ray bursts,  
$\rho_{\rm ND}^{\rm (AC)}=2.8 - 6.1\times 10^{11}~\mdens$.

The accreted crust is, in contrast to the GS one, a reservoir of the nuclear energy. 
This energy can be steadily released mainly at some 300 - 500 m  below the NS surface, during the accretion phase, 
leading to {\it deep crustal heating} \citep{Haensel1990,BBR1998}. The EOS of an AC is stiffer than that for the GS crust, particularly 
for densities $5\times 10^{11} - 5\times 10^{12}\;\mdens$. Consequently, the thickness of an AC is larger \citep{Zdunik2011}. 

In the present paper we present an approximate description to the NS crust structure in terms of the function  
relating the chemical potential and the pressure. Within  the one-component plasma model, we derive in 
Sect.\,\ref{sect:crust.approx} a formula for the thickness of any layer of the crust. It is highly accurate 
and does not require any knowledge of the EOS but the values of the chemical potential at the boundaries of a given layer in the crust. 
The approach is then extended to describe a GS crust and formulas for the NS radius, the crust thickness 
and mass depending only on the mass and radius  of the NS core are obtained. Their accuracy and the dependence on the choice of the
location of the core-crust transition are studied in Sect.\,\ref{sect:application}. In particular it is shown that the radius and mass 
of  a neutron star,
the crust thickness and its mass can be determined with an error smaller than 0.1\%, 0.3\%, 1\,\% and 5\% respectively.
In Sect.\,\ref{sect:acc_cat} the approximated approach is extended to the case of an accreted crust 
and a simple formula for the difference in the thickness of the AC and GS crusts is obtained. 
It only  involves  the total (integrated) energy release due to deep crustal heating,  
and the mass and radius of a NS with a GS crust and is extremely accurate ($< 1$~m).
Conclusions and perspectives are presented in Sect.\,\ref{sect:conclusion}.

\section{Crust structure: an approximation}
\label{sect:crust.approx}
The approximate approach to the macroscopic properties of the NS crust based
on the separation of the TOV equation into two factors dependent on stellar properties
(mass and radius) and the EOS of dense matter was discussed in \cite{Lattimer2007,Zdunik2002,Zdunik2008,Zdunik2011}.

The Tolman-Oppenheimer-Volkoff (TOV) equation of hydrostatic equilibrium in General Relativity is:
\begin{equation}
\frac{{\rm d}P}{{\rm d}r}=-\left(\rho+\frac{P}{c^2}\right)\left(1-\frac{2Gm}{rc^2}\right)^{-1}\left(\frac{Gm}{r^2}+4\pi Gr\frac{P}{c^2}\right)
\label{eqn:TOV}
\end{equation}
with $m=m(r)$ the gravitational mass enclosed in a sphere of radius $r$, $P$ the pressure and $\rho$ the mass-energy density.

The mass of the crust $M_{\rm crust}$ being small compared to the total mass $M$ of the NS, 
within the crust $m\approx M$
and $4\pi r^3P/mc^2\ll 1$. Consequently Eq.\,(\ref{eqn:TOV}) can be rewritten, in the 
crust\footnote{The term $4\pi r^3P/mc^2$ is of the order of $P/\rho c^2$ at the bottom of the inner crust 
but is  three orders of magnitude smaller than $P/\rho c^2$ at the neutron drip point. This is the reason for  keeping the factor $1+ P/\rho c^2$ while neglecting  the term  $4\pi r^3 P/mc^2$ as compared to one.}:
\begin{equation}
\frac{{\rm d}P}{ \rho+P/c^2}=
-GM \frac{\drom r }{ r^2 (1-{2GM/rc^2})}~.
\label{tovcr2}
\end{equation}
Let $P_{\rm cc}$, $n_{\rm cc}$ and $\mu_{\rm cc}=\mu(P_{\rm cc})$ be the pressure, baryon density and chemical potential at the core-crust interface, respectively.
Within the crust, i.e.
for $0<P<P_{\rm cc}$, we define a
dimensionless function of the local pressure:
\begin{equation}
\chi(P)=\int_0^{P} \frac{{\drom P^\prime}}{{\rho(P^\prime)c^2 + P^\prime}}~.
\label{chio}
\end{equation}

Notice that $\chi(P)$ is determined
solely by the EOS of the crust.
The integral of the right-hand side of Eq.\,(\ref{tovcr2}) then becomes:
\begin{equation}
\chi\left[P\left(r\right)\right]=\frac{1}{2}\ln
\left[\frac{1-r_{\rm g}/R}
{1-r_{\rm g}/r}\right]~,
\label{rap2}
\end{equation}
where $r_{\rm g}\equiv 2GM/c^2$. Defining $a=1-r_{\rm
g}/R$, we obtain $r$ within the crust as a function of
$\chi$:
\begin{equation}
r=r_{\rm g}/\left(1-a {\rm e}^{-2\chi}\right)~.
\label{eq:r-chi}
\end{equation}
In thermodynamic equilibrium  one can define the baryon chemical potential
$\mu={\drom \rho}/{\drom n}$. Thus, the first law of thermodynamics at $T=0$ implies:
\begin{equation}
 \mu=\frac{P+\rho c^2}{n}
 \label{eq:mu}
\end{equation}
which leads to the relation
\begin{equation}
 \frac{\drom P}{\rho c^2 + P}=\frac{\drom P}{\drom \mu} \frac{\drom \mu}{\rho c^2 + P}=\frac{\drom \mu}{\mu}~.
 \label{eq:dpdmu}
\end{equation}
The function $\chi$ is then given by $\exp(\chi)=\mu(P)/\mu_0$ where $\mu_0=\mu(P=0)=m_0c^2$  is the energy per 
baryon at NS surface.  

This allows to determine the thickness of any shell of the crust located between two radii $r_1$ and $r_2$ 
corresponding to the pressure $P_1$ and $P_2$, respectively:
\begin{equation}
  \frac{\sqrt{1-{2GM}/{r_1 c^2}}}{\sqrt{1-{2GM}/{r_2 c^2}}}=\exp{(\chi_{1,2})}{=}
\frac{\mu_2}{\mu_1}~.
\label{eq:shelthick}
 \end{equation}

A similar approach presented by 
\cite{Lattimer2007} relies on the replacement of $\mu$ in the denominator of Eq.\,(\ref{eq:dpdmu}) by its value at the NS surface
$\mu_0$ leading to an exponential dependence in Eq.\,(\ref{eq:shelthick}).

\subsection{Approximate formula for the radius and crust thickness}

Let $R_{\rm core}$ be the radius of the core, i.e. at the core-crust interface where $\mu=\mu_{\rm cc}$.
In Eq.\,(\ref{eq:shelthick}), taking $r_1=R$ and $r_2=R_{\rm core}$, one can obtain a formula relating $R(M)$ to $R_{\rm core}(M)$:
 \begin{equation}
{\frac{\sqrt{1-{2GM}/{Rc^2}}}{\sqrt{1-{2GM}/{R_{\rm core}c^2}}}=\frac{\mu_{\rm cc}}{\mu_0}}~.
\label{mrcore1}
\end{equation}

The latter is equivalent to:
 \begin{equation}
\frac{2GM}{Rc^2}=\frac{2GM}{R_{\rm core}c^2}-
\left(\frac{\mu_{\rm cc}^2}{\mu_0^2}-1\right)\left(1-\frac{2GM}{R_{\rm core}c^2}\right)~.
\label{mrcore1a}
\end{equation}
which expresses the compactness of the whole star in terms of the core compactness.
Note that in this formula the EOS of the crust enters through the ratio $\mu_{\rm cc}/\mu_0$.

From Eq.\,(\ref{mrcore1a}) we find that the radius is given by:
\begin{equation}
R=\frac{R_{\rm core}}{1-(\alpha-1)({R_{\rm core}c^2}/{2GM}-1)}
\label{mrcore2}
\end{equation}

where:
\begin{equation}
\alpha=\exp{(2\chi)}=
\left(\frac{\mu_{\rm cc}}{\mu_0}\right)^2~.
\label{alpha}
\end{equation}

Then the crust thickness $l_{\rm crust}=R-R_{\rm core}$ is:
\begin{equation}
l_{\rm crust}=\phi R_{\rm core}\frac{1-{2GM}/{R_{\rm core}c^2}}{1-\phi(1-{2GM}/{R_{\rm core}c^2})}
 \label{eq:thcr}
\end{equation}
where
\begin{equation}
\phi\equiv \frac{(\alpha-1)\rcore c^2}{2GM}~. 
 \label{eq:phidef}
\end{equation}

It should be noted that $\phi$ defined by Eq.\,(\ref{eq:phidef}) is a nonrelativistic quantity and 
can be approximated for $\alpha \to 1$ by:
\begin{equation}
\phi\simeq \frac{\delta\mu\rcore}{G M m_0}~~~~~~~~~~~\delta\mu=\mu_{\rm cc}-\mu_0
 \label{eq:phiap}
\end{equation}
The numerical formula  writes then:
\begin{equation}
 \phi\simeq 7.27 \times 10^{-3}\;\left(\frac{\delta\mu}{\rm MeV}\right)
 \left(\frac{R_{\rm core}}{\rm 10\ km}\right) \left(\frac{M}{\msun}\right)^{-1}~.
\label{eq:phiapnum}     
\end{equation}

The leading term in the expansion of the right-hand-side of Eq.\,(\ref{eq:thcr}) in 
powers of the parameter $\phi$  gives an approximate formula for the thickness
of the crust proportional to $\delta\mu$:
\begin{eqnarray}
l_{\rm crust}&\simeq& 1.82\,{\rm km}\cdot \left(1-\frac{2GM}{\rcore c^2}\right)
\left(\frac{\delta\mu}{\rm 25\,MeV}\right)\nonumber\\
&& \times \left(\frac{R_{\rm core}}{\rm 10\ km}\right)^2 \left(\frac{M}{\msun}\right)^{-1}
\label{eq:thicapnum}
\end{eqnarray}
where $\delta\mu$ is normalized to the ``typical'' value for the NS crust ($\sim 25$~MeV, see Table\,\ref{tab:cctrans}).
It should be however mentioned that for astrophysically relevant NS parameters ($M\sim 1-2\;\msun$,
$R_{\rm core}\sim 10-12$~km) one gets 
$\phi$ of the order of $0.15-0.25$. 
Therefore, $\phi$ cannot be considered  as a very small number. 
Thus the accuracy of the expansion given by Eq.\,(\ref{eq:thicapnum}) 
 is $\sim 20\%$ (see Fig.\,\ref{fig:thmmdhcr}) and one should instead use the formula in Eq.\,(\ref{eq:thcr}) 
to determine the thickness of the crust with a high accuracy ($<1\%$).

\subsection{Approximate formula for the mass of the neutron-star crust}

The crust contributes to the total mass of a NS. However the role
of the mass of the crust for the total stellar mass is by one order of magnitude smaller than the
importance of the crust thickness for the radius of a NS. To estimate the total mass of a NS with an accuracy similar to the one obtained using the approximation in Eq.\,(\ref{mrcore2}) for the radius, we can safely use a very crude approximation for the crust mass:
\begin{equation}
{\frac{\drom P}{\drom m}}=
- \frac{GM}{4\pi r^4 (1-{2GM/rc^2})}
\label{eq:dpdm}
\end{equation}
obtained from the TOV equation  by neglecting the $P/c^2$ terms.

The mass of the crust is given by the formula:
\begin{equation}
M_{\rm crust}=\frac{4 \pi P_{\rm cc} R_{\rm core}^4}{GM_{\rm core}}\left(1-\frac{2GM_{\rm core}}{R_{\rm core}c^2}\right)
\label{eq:mcrust}
\end{equation}
and is proportional to the pressure at the bottom of the crust $P_{\rm cc}$.
The total mass of the star is then $M=M_{\rm crust}+M_{\rm core}$.

Numerically:
\begin{eqnarray}
 {M_{\rm crust}}&\simeq& 7.62\times 10^{-2}\,\msun\cdot 
\left(\frac{P_{\rm cc}}{\rm MeV\,fm^{-3}}\right)\nonumber\\
& \times&\left(1-\frac{2GM_{\rm core}}{\rcore c^2}\right) \left(\frac{R_{\rm core}}{\rm 10\ km}\right)^4 \left(\frac{M_{\rm core}}{\msun}\right)^{-1}.
\label{eq:mcrapnum}
\end{eqnarray}

\section{Neutron-star parameters for a catalyzed crust}
\label{sect:application}
\subsection{Mass and radius of a neutron star from $\mu_{\rm cc}/\mu_0$}

\begin{figure}
\resizebox{\hsize}{!}{\includegraphics{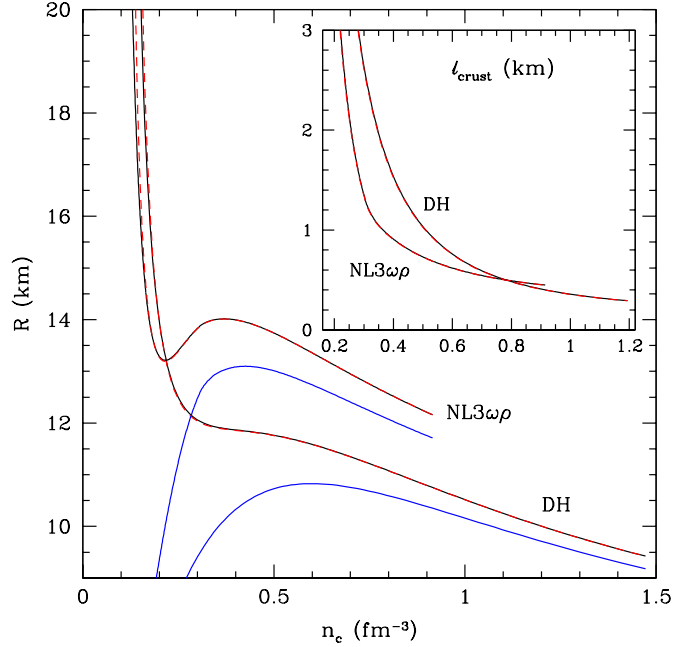}}
\caption{NS radius $R$ and thickness of the crust $l_{\rm crust}$ (inset) for the DH and NL3$\omega\rho$ EOS
as a function of the baryon density $n_{\rm c}$ at the center of the star.
Solid (black) curves - exact solution calculated for the unified EOS (i.e. including the crust EOS),
blue curves - radius of the core (above $P_{\rm cc}$),
dashed (red) lines - approximation based on Eq.\,(\ref{mrcore2}), obtained using the core EOS only.}
\label{fig:rnccra}
\end{figure}

\begin{figure}
\resizebox{\hsize}{!}{\includegraphics{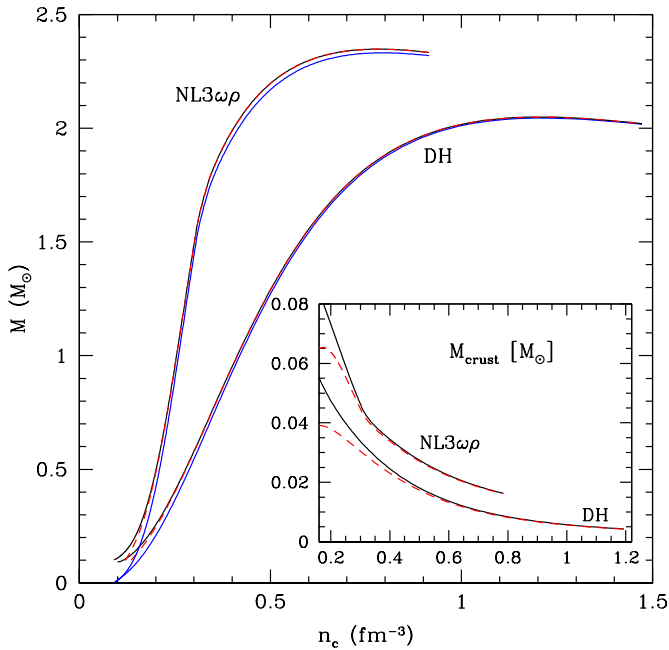}}
\caption{NS mass $M$ and mass of the crust $M_{\rm crust}$ (inset) for the DH and NL3$\omega\rho$ EOS
as a function of the baryon central density $n_{\rm c}$.
Solid (black) curves - exact solution calculated for the unified EOS (including the crust EOS),
blue curves - mass of the core ($P> P_{\rm cc}$),
dashed (red) lines - approximation based on Eq.\,(\ref{eq:mcrust}), obtained using the core EOS only.}
\label{fig:mnccra}
\end{figure}

\begin{table}
\begin{center}
\caption{Crust/core boundary for the two considered EOS DH and NL3$\omega\rho$ - the crust parameters for these EOSs are
presented in Figs.\,\ref{fig:rnccra}-\ref{fig:mrcr}. The bottom part of the table presents the artificial locations
of the crust/core boundary used to test the accuracy of the approximate approach in Figs.\,\ref{fig:mupcr},\,\ref{fig:mrdhcr}.}
\begin{tabular}{c|ccc}

 EOS & $n_{\rm cc}$ [fm$^{-3}$] & $P_{\rm cc}$ [MeV fm$^{-3}$]& $\mu_{\rm cc}$ [MeV] \\
\hline
\hline

&\multicolumn{3}{c}{real crust/core location}\\
\hline
&&&\\
DH     & 0.077 &  0.335 & 953.3    \\&&&\\
NL3$\omega\rho$     & 0.084 &  0.522 & 954.6  \\&&&\\
\hline
&\multicolumn{3}{c}{artificial crust/core location}\\
\hline
&&&\\

DH1     & 0.09 &  0.477 & 955.0    \\&&&\\
DH2     & 0.11 &  0.793 & 958.2    \\&&&\\
DH3     & 0.13 &  1.245 & 961.9    \\&&&\\
DH4     & 0.16 &  2.243 & 968.8    \\&&&\\
\end{tabular}
\label{tab:cctrans}
\end{center}
\end{table}

The approximate formulas presented in the previous section allow us to determine the main parameters of
a NS (total mass, radius, crust thickness and mass) on the basis of the properties  of its core only,
 i.e. using only an EOS $P(\rho)$ for nuclear matter below the crust/core
 interface (for $P>P_{\rm cc}$). 
The only additional information required is the chemical potential at zero pressure 
$\mu_0$. For cold catalysed matter the minimum energy is obtained for iron $^{56}$Fe:
 $m_0c^2=\mu_0=930.4$\,MeV \citep{NSBook}.

First, for a given central density $n_{\rm c}$ (or equivalently pressure $P_{\rm c}$, given by the core EOS),
and a chosen location of the core-crust transition at a density $n_{\rm cc}$ or pressure $P_{\rm cc}$, the relation between the 
mass and radius of the NS core $M_{\rm core}(R_{\rm core})$ is obtained by integrating the TOV equations outwards
from the center of a star with $P=P_{\rm c}$ down to $P_{\rm cc}$. Then the mass of the crust $M_{\rm crust}$ is
 determined using Eq.\,(\ref{eq:mcrust}) and consequently so is the total mass of the star $M$.  Using Eq.\,~(\ref{mrcore2})
the canonical $M(R)$ relation between the mass and the radius of the NS is reconstructed. The thickness of the crust is
finally given by Eq.\,(\ref{eq:thcr}).

\begin{figure}
\resizebox{\hsize}{!}{\includegraphics{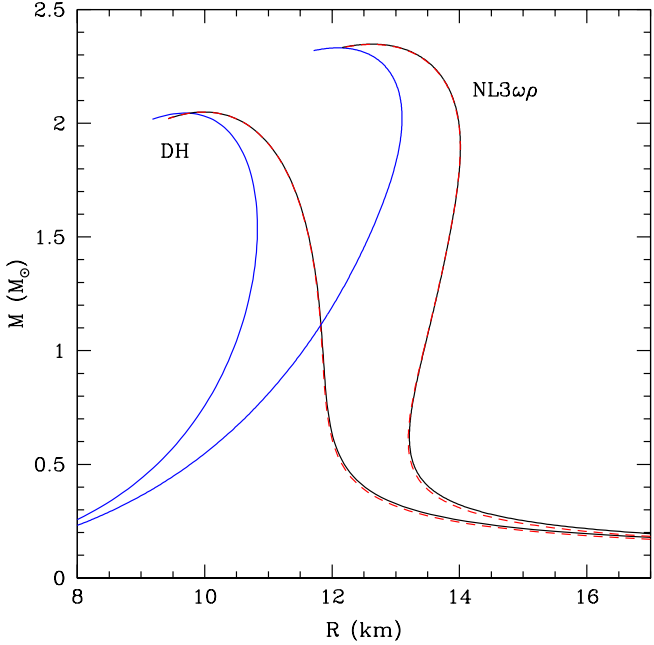}}
\caption{$M(R)$ dependence for the exact solution of the TOV equations obtained using a unified EOS and for our approximate  approach, for the DH and NL3$\omega\rho$ models
of dense matter.
Solid (black) curves - $M(R)$ calculated with a unified EOS,  
solid (blue) curves - $M_{\rm core}(R_{\rm core})$ relation,
dashed (red) lines - $M(R)$ approximation based on Eq.\,(\ref{mrcore2})-(\ref{alpha}),
 and the black ones -  the  exact solution of the TOV equation.}
\label{fig:mrcr}
\end{figure}

In Figs.\,\ref{fig:rnccra}, \ref{fig:mnccra} we present the result of a such procedure for two 
models of dense matter fulfilling  the observational constraint on the maximum allowable 
mass  $M_{\rm max }> 2\,\msun$: DH \citep{DH} and the stiffer NL3$\omega\rho$ \citep{Fortin2016}.
The approximate solution (dashed lines) is almost undistinguishable from the exact one except in the region of 
relatively small central density. Similar conclusions are obtained for the parameters of the crust (thickness and mass) presented in the insets.
In Fig.\,\ref{fig:mrcr} various relations between masses  and radii are presented.
The black solid lines are the $M(R)$ relations obtained when solving the TOV equations in the whole NS 
(core and crust)
with a unified EOS, i.e. when the same nuclear models for the crust and the core are used.
The blue solid lines correspond to the dependence $M_{\rm core}(R_{\rm core})$ 
and are obtained solving the TOV equations in the core i.e., 
from the center at pressure $P_{\rm c}$  outwards to the pressure at core/crust
interface $P_{\rm cc}$.
The red dashed lines are obtained using the $M(R_{\rm core})$ relations in Eq.\,(\ref{mrcore2}).
In the latter case, we do not need any information about the crust EOS
except the chemical potentials at zero pressure $\mu_0$ and at the bottom of the crust 
$\mu_{\rm cc}$ in order to determine the total radius of the star. 
For a chosen value of the density at the core-crust transition (see Sect.\,\ref{sect:choice}) 
$\mu_{\rm cc}$ can be calculated from Eq.\,(\ref{eq:mu}) using the core EOS.

The approximate formula Eq.\,(\ref{mrcore2}) works very well for astrophysically interesting masses of NS: $M>1\;\msun$. 
 For  the sake of completeness, let us mention that for masses as small as  $0.2\;\msun$ the validity of the 
formula breaks down, because the condition $M_{\rm crust} \ll M$ is then
obviously not fulfilled.
For  the NL3$\omega\rho$ EOS the difference between the exact and approximate radii is 
  20 m ($1.0\;\msun$), 8 m ($1.5\;\msun$) and 3 m ($2.0\;\msun$). Therefore, for $M>1\;\msun$
the relative error is less than 0.15\% of the radius of a star (or less than 1\% of the thickness of the crust). 
For the DH EOS the accuracy of
the approximate approach is even better.  For the estimation of the mass of the crust we use the simplest
approximations in Eq.\,(\ref{eq:mcrapnum}) which is accurate up to 6\% for $M>1\;\msun$ therefore resulting in a very small error in
the total mass determination (less than 0.3\%).

\subsection{Choice of the core-crust transition}
\label{sect:choice}

\begin{figure}
\resizebox{\hsize}{!}{\includegraphics[]{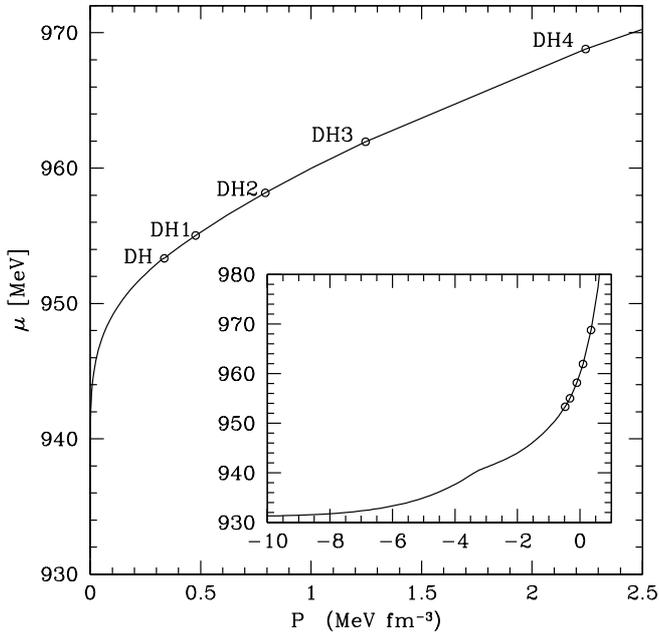}}
\caption{Baryon chemical potential $\mu$ as a function of the pressure for the DH EOS. 
The points mark the value of the chemical potential at various assumed  locations for the core-crust boundary used in Eq.\,(\ref{mrcore2})-(\ref{alpha})
and presented in Table \ref{tab:cctrans}. The lowest point (DH) corresponds to the real core-crust
interface in the DH model. Inset: $\mu(P)$ in logscale for the pressure.}
\label{fig:mupcr}
\end{figure}

At the core-crust interface the ground state of neutron star matter changes from a lattice of spherical nuclei in the solid crust to homogeneous matter in the liquid core. Some models predict the appearance of so-called pasta phases when the most stable shape of nuclei is not any more a sphere but, as the density increases, a rod or a slab immersed in the neutron gas \citep{RP83}.
Various approaches have been developed to determine the density of the core-crust transition $n_{\rm cc}$ eg. the study of thermodynamic spinodal or dynamical spinodal surfaces, Thomas Fermi calculations or the Random Phase Approximation. However for $\beta$-equilibrated matter, the values of $n_{\rm cc}$ that are obtained have been shown to be similar (see for example \cite{HS08,AC10a,AC10b,Pais2016} and references therein).

The transition density is inversely proportional to the slope of the symmetry energy $L$ \citep{HP01} and for typical values of $30\leq L \leq 120$ MeV, \cite{DM11} find for a large set of EOS based on two nuclear approaches to the many-body problem (Skyrme models and Relativistic Mean Field calculations)
$0.06 \leq n_{\rm cc} \leq 0.10$~fm$^{-3}$ or $0.38 \leq n_{\rm cc}/n_0 \leq 0.63$. 
Consequently, if for a given core EOS no calculation of the core-crust transition density is available, 
taking $n_{\rm cc}=0.5 n_0$ appears reasonable.

In this subsection we discuss the accuracy of our approximate approach to calculate the main parameters of a NS
for different given locations of the crust-core boundary, using the DH model of dense matter as an example. 
In this model the `real' crust-core boundary is located at $n_{\rm cc}=0.077\,{\rm fm}^{-3}$,
i.e. at about half nuclear matter density. To test the dependence of our approximations on  $n_{\rm cc}$ we
artificially re-define the location of the core boundary to 
 $0.09,\,0.11,\,0.13,\,0.16\,{\rm fm}^{-3}$ (models DH1-DH4 in Table \ref{tab:cctrans}).
 The size of the core, which is defined by the pressure $P_{\rm cc}$, decreases with increasing $P_{\rm cc}$ and is
 smallest (at given central pressure $P_{\rm c}$) for the DH4 model. Consequently the region of the star described by the
 approximate formulas (outer part $P<P_{\rm cc}$) is larger for larger  $P_{\rm cc}$ and for the DH4 model 
 (with $n_{\rm cc}\simeq n_0$) the mass of the crust is about $0.1\,\msun$, which is much larger than the mass of the real crust (unified DH EOS, $<0.02\,\msun$).

Fig.\,\ref{fig:mupcr} shows the baryon chemical potential $\mu$ as a
function of the pressure $P$ for the DH EOS of the core. The dots correspond to the considered crust/core
location (see Table \ref{tab:cctrans}). The lowest value is the real density of the crust-core
interface for the (unified) DH model. 

\begin{figure}
 
\resizebox{\hsize}{!}{\includegraphics{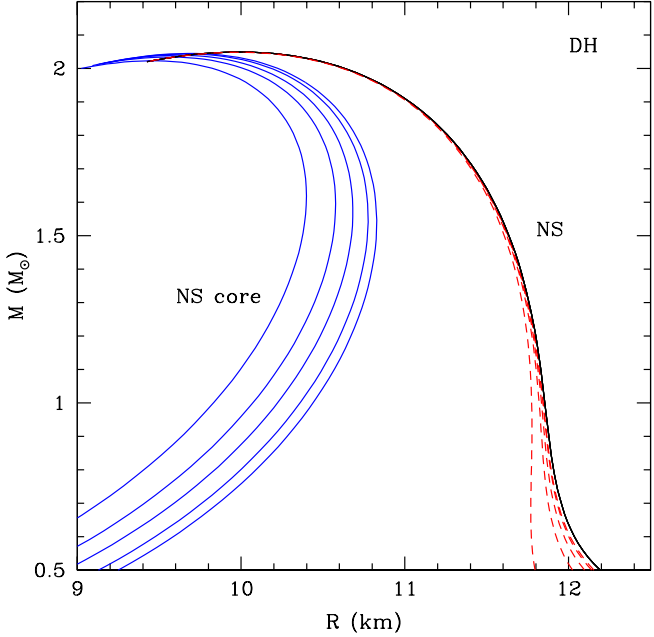}}
\caption{$M(R)$ dependence for the exact solution of the TOV equations obtained using a unified EOS and the approximate approach for the DH EOS. 
Different values of $n_{\rm cc}$ are used (from left to right; $0.16,\,0.13,\,0.11,\,0.09,\,0.077\,{\rm fm}^{-3}$, see 
Table\,\ref{tab:cctrans} for details). 
The value $n_{\rm cc}=0.077\,{\rm fm}^{-3}$ corresponds to the `real' crust/core boundary for the DH model.
The solid blue curves show the $M(R_{\rm core})$ relation, the dashed red ones the $M(R)$ approximation based on Eq.\,(\ref{mrcore2})-(\ref{alpha})
 and the black one the exact solution of the TOV equation.}
\label{fig:mrdhcr}
\end{figure}

In Fig.\,\ref{fig:mrdhcr} we present the results of our approximate approach given by Eq.\,(\ref{mrcore2})-(\ref{alpha}) for the mass-radius relation for the DH EOS.
For the real value of $n_{\rm cc}$ the difference 
between the exact and approximate results is 30 m at $M=0.5\,\msun$, 10 m at $M=1\,\msun$ and 
less than 4 m at $M=1.5 - 2\, \msun$. Even for an unrealistically large value of $n_{\rm cc}= n_0=0.16\,{\rm fm}^{-3}$ 
(DH4) the approximation gives a quite high accuracy with the uncertainty on $R$ decreasing from $\sim100$ m at $1\,\msun$ down to
30 m at $1.5\,\msun$ and 5 m at $2\,\msun$.

\begin{figure}
\resizebox{\hsize}{!}{\includegraphics{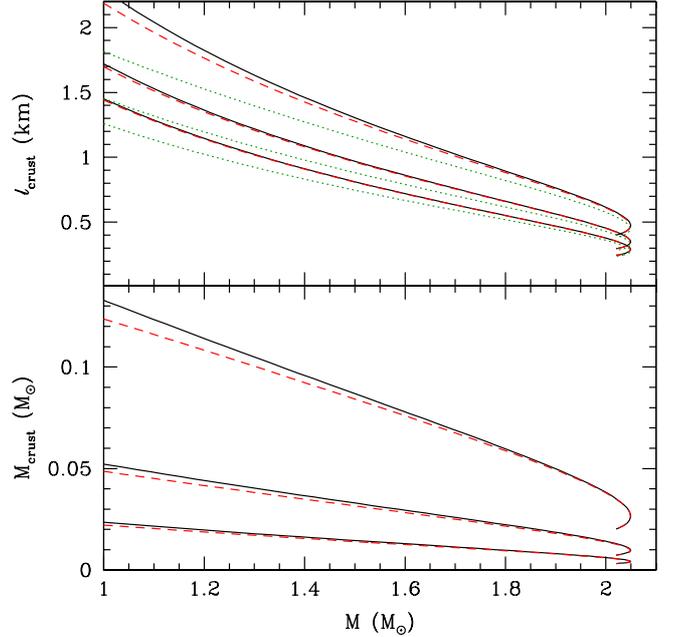}}
\caption{Thickness $l_{\rm crust}$ (upper panel) and mass $M_{\rm crust}$ (lower panel) of the crust of a NS for the DH EOS 
and for different values of $n_{\rm cc}$ (from top to bottom; $0.16,\,0.11,\,0.077\,{\rm fm}^{-3}$). 
The density $n_{\rm cc}=0.077\,{\rm fm}^{-3}$ corresponds to the crust-core boundary for the DH model.
Solid lines - exact results calculated for the complete EOS (including the crust EOS),
dashed (red) lines -  approximations based on Eq.\,(\ref{eq:thcr}) (thickness) and  Eq.\,(\ref{eq:mcrust}) (mass), 
dotted (green) lines - linearisation Eq.\,(\ref{eq:thicapnum}) of Eq.\,(\ref{eq:thcr}).}
\label{fig:thmmdhcr}
\end{figure}

The accuracy of the approximate approach for the thickness and mass of the crust is presented in Fig.\,\ref{fig:thmmdhcr}
for three different locations of the bottom of the crust (models DH, DH2, DH4). 
The thickness of the crust is determined very accurately by the formula (\ref{eq:thcr}).
The relative error is less than $0.7\%$ for DH and $<3.5\%$ for DH4 which is equivalent to 
$<10$~m and $<100$~m inaccuracy 
(i.e. $0.08\%$, $0.7\%$ error in the radius of the star $R$).
Although for the presented range of masses from
$1\,\msun$ to $M_{\rm max}$
the maximum relative error in $M_{\rm crust}$ is 5-8\%, the determination of the total mass $M$ of neutron star
 is very accurate (1\% for $n_{\rm cc}=n_0$ and 0.1\% for $n_{\rm cc}=0.5n_0$ at $M=1\,\msun$, the error decreasing 
rapidly with further increase of $M$).

\subsection{Accuracy of the approximate approach and crust/core matching problem}

In the case of a non-unified EOS the matching of the core EOS to the crust EOS is often performed by 
an artificial function $P(\rho)$ or $P(n)$ (which can be linear, polytropic, \ldots). In general this approach
leads to thermodynamic inconsistency, which manifests itself in a discontinuity in $\mu$ 
(for details see \citealt{Fortin2016}). This
discontinuity  $\delta\mu_{_{\rm EOS}}$ can be as high as few MeV, 
but is usually of the order of $\delta\mu_{_{\rm EOS}}\simeq 0.5-1.5$~MeV.
It results in an error in the crust thickness determination which can be calculated with Eq.\,(\ref{eq:thicapnum}). 
The relative error in $l_{\rm crust}$ is 
then proportional to $\delta\mu_{_{\rm EOS}}/\Delta\mu$ where $\Delta\mu$ is the chemical potential range in the crust (20-30 MeV).
For example, for $\delta\mu_{_{\rm EOS}}\simeq 1$~MeV the error due to inconsistent crust/core matching is 
larger than the accuracy given by our approximation for the crust thickness (for $n_{\rm cc}=0.5\;n_0$ and $M=1.4\,\msun$
it is 40~m. compared to 5~m inaccuracy of our model). 

As a consequence, using the approximate formula for the radius: Eq.\,(\ref{mrcore2}) and the crust thickness: Eq.\,(\ref{eq:thcr}) and mass: Eq.\,(\ref{eq:mcrust}), without any further knowledge about the crust EOS, is in general 
more accurate than the widely-used method of matching in a thermodynamically inconsistent way an EOS for the crust to a different (non-unified) one for the core.

\section{Accreted vs catalyzed crust}
\label{sect:acc_cat}

\begin{figure}
\resizebox{\hsize}{!}{\includegraphics{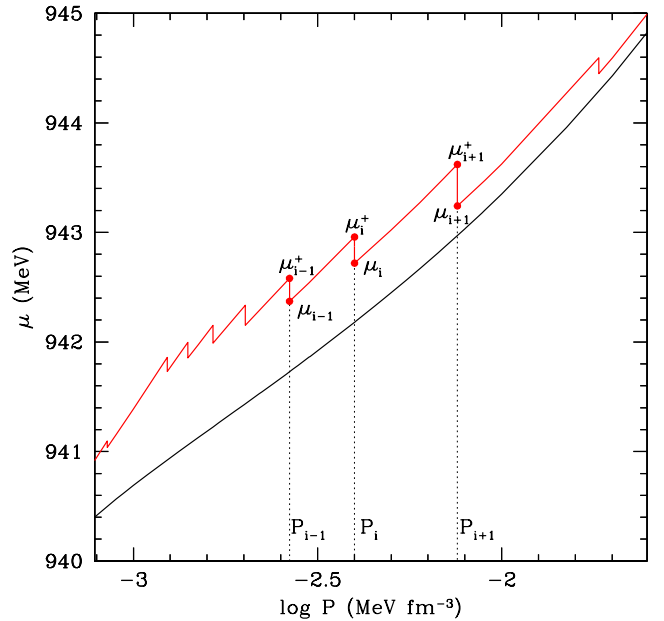}}
\caption{Baryon chemical potential $\mu$ for catalyzed (lower curve) and accreted crust (step-like curve).
The \cite{Mackie1997} model of dense matter is used in the example (for details see  \cite{Haensel2008}).} 
\label{fig:mupla}
\end{figure}

A characteristic feature of the EOS for an accreted crust is the existence of energy sources at pressures at which exothermic nuclear reactions are induced by the accretion of matter onto the NS surface. As a result the $\mu(P)$ relation is a discontinuous function with drops in $\mu$ equal to the
energy release per one  accreted nucleon. In Fig.\,\ref{fig:mupla} we present an example 
for  the \cite{Mackie1997} model of dense matter \citep{Haensel2008} of a continuous $\mu^{^{\rm (GS)}}(P)$ dependence
for catalyzed matter (lower continuous curve which corresponds to the minimum value of $\mu$ at a given pressure) and for an accreted crust (with energy sources
located at $P=P_{i-1}, P_i, P_{i+1}$).

\subsection{Thickness of an accreted crust}

In Sect.\,\ref{sect:crust.approx} we considered a catalysed crust for which the function $\mu(P)$  is continuous
 and Eqs.\,(\ref{eq:mu},\ref{eq:dpdmu}) hold. In this case the formula (\ref{eq:shelthick}) can be used for the whole crust,
resulting in Eq.~(\ref{eq:thcr}).
In the case of an accreted crust, $\mu(P)$ is not continuous as shown in Fig.\,\ref{fig:mupla} and each jump in the chemical potential at a fixed pressure corresponds to an energy source. Consequently, in order to determine the thickness of an accreted crust one has to calculate separately
the thickness of each shell located between two energy sources, for example between $P_{i-1}$ and $P_i$
as plotted in Fig.\,\ref{fig:mupla}.

Using Eq.\,(\ref{eq:shelthick}) the following set of equations is obtained: 
 \begin{align*}
\frac{\sqrt{1-\frac{2GM}{Rc^2}}}{\sqrt{1-\frac{2GM}{R_1c^2}}}&=\exp{\chi_1}=
\frac{\mu_1^+}{\mu_0}\\
\frac{\sqrt{1-\frac{2GM}{R_1c^2}}}{\sqrt{1-\frac{2GM}{R_2c^2}}}&=\exp{\chi_2}=
\frac{\mu_2^+}{\mu_1}\\
\dots\\
\frac{\sqrt{1-\frac{2GM}{R_ic^2}}}{\sqrt{1-\frac{2GM}{R_{i+1}c^2}}}&=\exp{\chi_{i+1}}=
\frac{\mu_{i+1}^+}{\mu_i}\\
\dots \\
\frac{\sqrt{1-\frac{2GM}{R_n c^2}}}{\sqrt{1-\frac{2GM}{R_{\rm core}c^2}}}&=\exp{\chi_{n+1}}=
\frac{\mu_{\rm core}^+}{\mu_n}
\label{accshells}
\end{align*}
where the subscript `core' corresponds to the convergence point of the baryon chemical potential for
accreted and catalyzed crusts at the bottom of the crust, i.e. where the condition
$\mu_{\rm core}^+=\mu_{\rm core}$ is fulfilled (see Fig.\,\ref{fig:mupla}).

Multiplying the above equations by one another we get the final formula for the thickness 
of an accreted crust:
\begin{eqnarray}
 \frac{\sqrt{1-\frac{2GM}{Rc^2}}}{\sqrt{1-\frac{2GM}{R_{\rm core}c^2}}}&=&
 \frac{\mu_1^+}{\mu_1}\cdot  \frac{\mu_2^+}{\mu_2} \cdots \frac{\mu_{i}^+}{\mu_i}\cdots
 \frac{\mu_n^+}{\mu_n}\cdot \frac{\mu_{\rm core}}{\mu_0}\nonumber\\
 &=& \frac{\mu_{\rm core}}{\mu_0}\cdot\prod_{i=1}^{n}\frac{\mu_{i}^+}{\mu_i}
 \label{eq:acth}
\end{eqnarray}
where the product is calculated over all the energy sources in the accreted crust.

The energy release per one accreted nucleon at the  pressure $P_i$  is given by $Q_i=\mu_i^+-\mu_i$.
Because $Q_i/\mu_i < 10^{-3}$, one can safely approximate formula Eq.\,(\ref{eq:acth}) by 

\begin{equation}
  \frac{\sqrt{1-\frac{2GM}{Rc^2}}}{\sqrt{1-\frac{2GM}{R_{\rm core}c^2}}}\simeq
  \frac{\mu_{\rm core}}{\mu_0}(1+\sum_{i=1}^{n}\frac{Q_i}{\mu_i})~.
  \label{eq:acthap1}
\end{equation}

The main energy sources for an accreted crust are located in the inner crust at typical
pressures $P\sim 0.001-0.01\,{\rm MeV}\,{\rm fm^{-3}}$ ($10^{30}-10^{31}\,{\rm erg}\,{\rm cm^{-3}}$) 
where the chemical potential
$\mu_{\rm IC}\simeq 942\,{\rm MeV}$.
Replacing $\mu_i$ in Eq.\,(\ref{eq:acthap1}) by this typical value, we get:

\begin{equation}
  \frac{\sqrt{1-\frac{2GM}{Rc^2}}}{\sqrt{1-\frac{2GM}{R_{\rm core}c^2}}}\simeq
  \frac{\mu_{\rm core}}{\mu_0}
  \left(1+\frac{Q_{\rm tot}}{\mu_{\rm IC}}\right)~~~~~Q_{\rm tot}=\sum_{i=1}^{n}Q_i~,
  \label{eq:acthap2}
\end{equation}
where $Q_{\rm tot}$ is the total energy release in the crust.

\subsection{Thickness of a catalyzed crust vs. an accreted one}

The formulas for the radius $R_{\rm cat}$ and $R_{\rm acc}$ of a NS with a catalysed crust and an accreted one, respectively, are:
\begin{align}
\frac{\sqrt{1-\frac{2GM}{R_{\rm cat}c^2}}}{\sqrt{1-\frac{2GM}{R_{\rm core}c^2}}}&=\frac{\mu_{\rm core}}{\mu_0}~,
\label{eq:rcat}\\
 \frac{\sqrt{1-\frac{2GM}{R_{\rm acc}c^2}}}{\sqrt{1-\frac{2GM}{R_{\rm core}c^2}}}&=
 \frac{\mu_{\rm core}}{\mu_0}\cdot\prod_{i=1}^{n}\frac{\mu_{i}^+}{\mu_i}~.
 \label{eq:racc}
\end{align}
They are equivalent to:
\begin{equation}
  \frac{\sqrt{1-\frac{2GM}{R_{\rm acc}c^2}}}{\sqrt{1-\frac{2GM}{R_{\rm cat}c^2}}}=
  \prod_{i=1}^{n}\frac{\mu_{i}^+}{\mu_i}
  \simeq 1+\frac{Q_{\rm tot}}{\mu_{\rm IC}}~.
  \label{eq:rmuap}
\end{equation}

Defining $\sqrt{\alpha}\equiv \prod_{i=1}^{n}\frac{\mu_{i}^+}{\mu_i}$,
 Eq.\,(\ref{mrcore2}) holds with
$R$ and $R_{\rm core}$ replaced by $R_{\rm acc}$ and $R_{\rm cat}$ respectively, i.e.:
\begin{equation}
\frac{R_{\rm cat}}{R_{\rm acc}}=1-(\alpha-1)\left(\frac{R_{\rm cat}c^2}{2GM}-1\right)~.
\label{eq:raccat}
\end{equation}

The difference in the radii of  NS with an accreted crust and with a catalyzed one: $\Delta R= R_{\rm acc}-R_{\rm cat}$ is small compared to $R$ and one gets
the approximate relation:
\begin{equation}
 \frac{\Delta R}{R_{\rm cat}}\simeq \left[\left(\prod_{i=1}^{n}\frac{\mu_{i}^+}{\mu_i}\right)^2-1\right]
 \left(\frac{R_{\rm cat}c^2}{2GM}-1\right)~.
 \label{eq:raccatap}
\end{equation}

Using Eq.\,(\ref{eq:rmuap}) we obtain:
\begin{equation}
 \frac{\Delta R}{R_{\rm cat}}\simeq 
 2\left(\frac{R_{\rm cat}c^2}{2GM}-1\right)\sum_{i=1}^{n}\frac{Q_i}{\mu_i}
 \simeq 2\frac{Q_{\rm tot}}{\mu_{\rm IC}}\left(\frac{R_{\rm cat}c^2}{2GM}-1\right)
 \label{eq:raccatap1}
\end{equation}
which, after normalization to typical values, becomes:
\begin{equation}
 \Delta R\simeq 
144\,{\rm m}\cdot \left(\frac{Q_{\rm tot}}{2\,\rm MeV}\right)\left(\frac{ R_{\rm cat}}{\rm 10\, km}\right)^2
\left(\frac{M}{\msun}\right)^{-1}
\left(1-\frac{2GM}{R_{\rm cat}c^2}\right)~,
 \label{eq:raccatapnum1}
\end{equation}
where we used $\mu_{\rm IC}=942$~MeV. 
The inaccuracy of the formula (\ref{eq:raccatapnum1})  introduced 
by the approximations (\ref{eq:acthap1},\ref{eq:acthap2},\ref{eq:raccatap1}) 
is of the order of ($Q_{\rm tot}/940{\,\rm MeV}$), about 0.1\%   
in $\Delta R$ (i.e. much less than 1 meter).

\begin{figure}
\resizebox{\hsize}{!}{\includegraphics{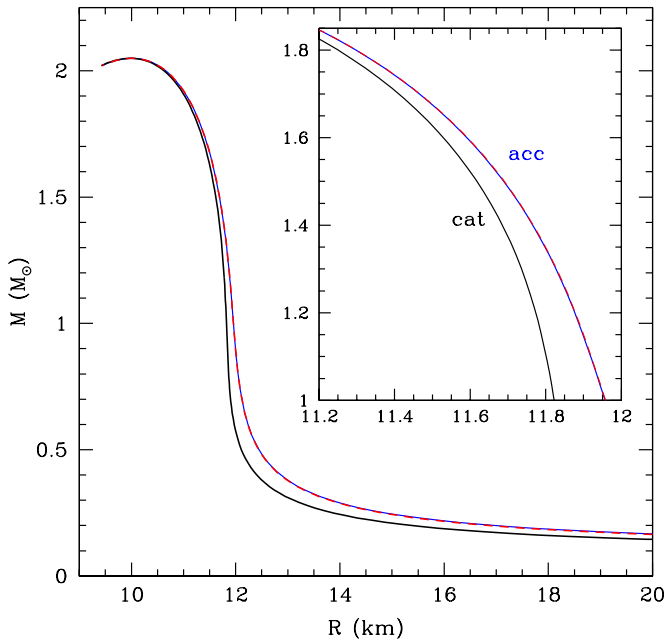}}
\caption{Mass-radius relation for the DH EOS for catalyzed (black, solid curve) and accreted crusts (blue, solid line).
The approximation given by Eq.\,(\ref{eq:raccatapnum1}) is plotted by a dashed red curve which can be hardly distinguished from the exact
result for an accreted crust. Inset - zoom for masses $1-2\,\msun$.}
\label{fig:mrapp}
\end{figure}

The accuracy of our approximation is visualized  in Fig.\,\ref{fig:mrapp}. The exact $M(R)$ curve is obtained for the
DH EOS with catalysed and accreted crusts based on \cite{Mackie1997} model of nuclei \citep{Zdunik2011,Haensel2008}. 
The difference in radii due to the formation scenario (ie. accreted vs.
catalyzed matter) is $\Delta R = 80$~m for $1.4\,\msun$ and in the range $330-50\,{\rm m}$ for masses between
$0.5\,\msun$ and $1.8\,\msun$ respectively. For the considered model of accreted matter the total energy release is
$Q_{\rm tot}=1.9\,{\rm MeV}$ and the $M(R)$ dependence obtained using Eq.\,(\ref{eq:raccatapnum1}) nearly
coincides with the exact one, with a difference of about 1~m for $M=1.5\,\msun$.

Let us mentioned that when deriving Eqs.\,~(\ref{eq:rcat},\ref{eq:racc}) 
we assume the same value of $\muco$ and $\mu_0$ for both an accreted crust and a catalyzed one. For $\muco$ this assumption is justified; indeed the
baryon chemical potential $\mu$ for catalyzed and accreted crusts converges at pressures larger than $\sim 0.03\,{\rm MeV fm^{-3}}$ (see Fig.\,\ref{fig:mupla}).
However for an accreted crust the value of $\mu_0$ actually depends on the ashes of nuclear reactions at the surface of the NS (as a result of the X-ray bursts,
\citealt{Haensel2003}).
In principle it is possible that $\mu_{0{\rm acc}}\ne \mu_{0{\rm cat}}$ and the product $\prod_{i=1}^{n}\frac{\mu_{i}^+}{\mu_i}$
should then be multiplied by $\mu_{0{\rm acc}}/\mu_{0{\rm cat}}$. In practice the difference is smaller than
$0.2\,{\rm MeV/nucleon}$ (0.02\%) for the ashes considered by \cite{Gupta2007,Haensel2008}.

\section{Conclusions}
\label{sect:conclusion}
In this paper we present an approximate treatment for the crust of a NS which allows to
calculate the mass and radius of a NS, the crust thickness and the crust mass. 
Two limiting  cases  were considered: catalyzed (ground-state) crust and fully accreted crust. 
For a catalyzed crust $R$, $l_{\rm crust}$, $M_{\rm crust}$, and $M$ do not depend on the specific form
of the crust EOS, but on the crust-core transition density only and on the EOS core. 
 For a given core EOS and chosen density at which the 
core-crust transition takes places, the relation between the core radius and mass is obtained by solving 
the TOV equations. Then the mass and radius of the crust and the total mass and radius of the star can be obtained using simple formulas.
The accuracy of this approach is higher than 1\% and 5\%  for  $l_{\rm crust}$ 
and $M_{\rm crust}$ respectively, for NS masses larger than $1\,\msun$. 
This is equivalent to the determination of global parameters of NS (radius $R$ and mass $M$) with maximum error
$\sim 0.1-0.3$\%.
Notice that unless if available in the literature, 
for a given EOS of the core, the transition density $n_{\rm cc}$ between the core and the crust is not known in advance. However for 
reasonable values of the symmetry energy usually $n_{\rm cc}\simeq 0.5\;n_0$. 
A simple and accurate formula 
for the difference in the radii of a NS with a fully accreted crust with respect to one with a catalyzed crust is derived. It is proportional to the total energy release due to deep crustal heating and depends 
in addition only on the mass and radius of the model with a catalyzed crust. 

The demonstrated high precision of the prediction of the radius of a NS makes the derived formulas of interest 
for theoretical works in particular in relation with the next generation of X-ray telescopes
which are expected to provide measurements of the NS radius with a precision of few percent.

\begin{acknowledgements}
This work was partially supported by the Polish National Science Centre (NCN)  grant no. 2013/11/B/ST9/04528 and the COST Action NewCompStar. MF would like to thank Constança Provid\^encia for useful discussions. We also acknowledge helpful remarks of participants of the MODE-SNR-PWN
2016 Workshop (Meudon, France), after the talk by one of the
authors (MF).
\end{acknowledgements}


\end{document}